\documentclass[superscriptaddress,twocolumn,showpacs,showkeys,amsmath]{revtex4}

\usepackage{graphicx} 

\def\eV{\hbox{ eV}}

\def\GeV{\hbox{ GeV}}

\def\mm{\hbox{ mm}}

\def\y{\hbox{ y}}

\begin{document}

\title{Extra Dimensions and Neutrinoless Double Beta Decay
Experiments}

\author{Marek G\'o\'zd\'z} 
\email{mgozdz@kft.umcs.lublin.pl}
\author{Wies{\l}aw A. Kami\'nski}
\email{kaminski@neuron.umcs.lublin.pl} 
\affiliation{Theoretical Physics Department, Maria
Curie--Sk{\l}odowska University, Lublin, Poland}

\author{Amand Faessler}
\email{amand.faessler@uni-tuebingen.de} 
\affiliation{Institute f\"ur
Theoretische Physik, Universit\"at T\"ubingen, Auf der Morgenstelle 14,
D-72076 T\"ubingen, Germany }

\begin{abstract}
The neutrinoless double beta decay is one of the few phenomena,
belonging to the non-standard physics, which is extensively being sought
for in experiments. In the present paper the link between the half-life
of the neutrinoless double beta decay and theories with large extra
dimensions is explored. The use of the sensitivities of currently
planned $0\nu2\beta$ experiments: DAMA, CANDLES, COBRA, DCBA, CAMEO,
GENIUS, GEM, MAJORANA, MOON, CUORE, EXO, and XMASS, gives the
possibility for a non-direct `experimental' verification of various
extra dimensional scenarios. We discuss also the results of the
Heidelberg--Moscow Collaboration. The calculations are based on the
Majorana neutrino mass generation mechanism in the
Arkani-Hamed--Dimopoulos--Dvali model.
\end{abstract}


\maketitle

\section{Introduction}

The standard model (SM), despite being very successful, has many
drawbacks and cannot be regarded as the ultimate theory of elementary
particles and their interactions. One of the most severe are the big
number of free parameters and the hierarchy problem. The idea of going
beyond the SM is old, but only recently has been fully backed up by
experimental evidence of neutrino oscillations
\cite{SNO,SNO2,SNO3,Ahn,s-kamiokande}. This observation clearly calls
for a more fundamental theory, which will extend the SM. One of the best
candidates, which cures most of the SM's weaknesses, is the theory of
superstrings. Within this theory one gets rid of the problematic
point-like nature of particles, therefore removing ultraviolet
divergences. What is more, there is no need to introduce and fit to
experiments so many free parameters (more than 20 in the SM). The string
theory is a consistent, divergent-free quantum field theory. It allows
also to generate, in a natural way, a spin-2 massless field, which
represents the graviton, therefore unifying gravity and electroweak
interactions.

However, the string theory requires two new features for consistent
formulation: the supersymmetry and additional spatial dimensions.
Supersymmetry introduces new kind of operators, which change the spin of
particles by 1/2. Therefore a boson may be turned into a fermion and
vice versa, which results in unification of forces and matter. The idea
of additional spatial dimensions sound even more strange. Because these
two phenomena are not observed in the low energy domain, the
supersymmetric particles must be heavier than our current experimental
capabilities, and extra dimensions much smaller than our ability to
observe them. They may form closed, curled shapes with tiny radii (the
so-called Calabi--Yau shapes) which has very small influence on our
life. It is, however, possible that the extra dimensions are open,
\emph{large}, and the difficulty in observing them has a different
basis.

There are three main models used in building theories with large extra
dimensions. Firstly, we have the description based on the original
Kaluza--Klein approach \cite{kaluza,klein}, which predicts a load of new
particles, the so-called towers of KK excitations. Secondly, there are
models formulated by Randall and Sundrum (RS models) \cite{rs1,rs2}, in
which the extra dimensions are compactified on orbifolds with a $Z_2$
symmetry. Last but not least, there are propositions of Arkani-Hammed,
Dimopoulos and Dvali (ADD models) \cite{add1,add2,add3,add4,add5}, in
which the standard model particles are trapped on a 3D brane, which in
turn floats in a higher dimensional bulk. In what follows we will use
the ADD approach.

As mentioned above, the main evidence for physics beyond the SM are the
oscillations of neutrinos. According to most theories, neutrino
oscillations imply a non-zero neutrino mass and a difference between the
flavor and mass eigenstates of neutrinos. The physical mechanism
generating this difference, as well as masses, remain still an open
issue.

Assuming the existence of supersymmetry, or non-standard physics in
general, one may expect the phenomena, normally forbidden by the
symmetries of SM. One of these is the neutrinoless channel of double
beta decay ($0\nu2\beta$) \cite{doi}. This decay, still not confirmed,
is the only experimental possibility of determining the nature of
neutrino (Majorana or Dirac particle). It may also help in obtaining the
absolute mass of these particles, since the oscillation experiments are
sensitive to differences of masses squared only.

In our earlier work \cite{mg-extradim1,mg-extradim2} we have shown, that
it is possible to relate the parameters describing extra dimensions to
the half-life of the $0\nu2\beta$ decay. In the present paper we
continue this topic with the analysis of the data provided by the
Heidelberg--Moscow Collaboration (H--M) as well as sensitivities of
planned $0\nu2\beta$ experiments, from the point of view of large extra
dimensions.

In the next section we recall the necessary information about the ADD
model of large extra dimensions. In the following section we shortly
discuss the formula describing the half-life of neutrinoless double beta
decay. After that, we arrive at the final formula and present and
discuss the results. A summary follows at the end.

\section{Extra dimensions}

In the ADD approach the space can be divided into two main parts. One
is the so-called bulk, a $3+n-$dimensional space in which the gravity
propagates. Besides gravitons, it may be also populated with other
particles and fields, which are not contained in the standard model.
Floating in the bulk, there is at least one 3-dimensional brane, an
object which is predicted by the string theory as a higher-dimensional
generalization of the string. The standard model is assumed to be
confined on such brane. It means that all the SM interactions are
propagating only within the brane. The same goes to fermions, which
must also be restricted to live on the brane. The just presented
setting immediately explains the weakness of gravity in our universe,
thus solving the hierarchy problem. Namely, using for example the
generalized Gauss Law, one arrives at the so-called reduction formula
\cite{add4}
\begin{equation}
  M_{Pl}^2 \sim R^n M_*^{2+n},
\label{rf}
\end{equation}
where $M_{Pl} \sim 10^{28} \eV$ is the Planck mass, $M_*$ is the true
scale of gravity, and $R$ is the assumed common compactification radius
of extra dimensions. One sees that, by properly adjusting the values of
$R$ and $n$, it is possible to lower the true scale of gravity to the
electroweak scale $\sim1$ TeV, thus getting rid of the hierarchy
problem. This is the main motivation for the ADD approach.

Introducing a second brane, parallel to ours, one may generate, under
special conditions, a Majorana neutrino mass term. Let us denote the
coordinates by $\{x^\mu,y^m\}$, where $\mu=0\dots 3$ labels the ordinary
space-time coordinates and $m=1 \dots n$ labels the extra dimensions.
By identifying $y\sim y+2\pi R$ we compactify the extra dimensions on
circles. (From now on we will drop the indices $\mu$ and $m$ for
simplicity.)

Let us assume \cite{add4, add5} that lepton number is conserved on our
brane, located at $y=0$, but maximally broken on the other one, placed
on $y=y_*$. The breaking occurs in a reaction where a particle $\chi$,
with lepton number $L=2$ and mass $m_\chi$, escapes the other brane into
the bulk. This particle, called the \emph{messenger}, may interact with
our brane and transmit to us the information about lepton number
breaking.

To be more specific, let us introduce a field $\phi_{L=2}$ located on
the other brane, whose vacuum expectation value (vev) breaks the lepton
number. What is more, it acts as a source for the bulk messenger field
$\chi$ and ``shines'' it everywhere, in particular also on our brane.
The strength of the shined $\chi$ is in a natural way suppressed by the
distance $r$ between branes, and therefore one can write for the
messenger
\begin{equation}
  \langle \chi \rangle = \langle \phi \rangle \Delta_n(r),
\label{chi}
\end{equation}
where $\Delta_n(R)$ is the $n$-dimensional propagator given by
\begin{eqnarray}
  \Delta_2(R) &\sim &
  \begin{cases}
    -\log(Rm_\chi), &(Rm_\chi \ll 1) \\
\frac{e^{-Rm_\chi}}{\sqrt{Rm_\chi}}, &(Rm_\chi \gg 1)
\end{cases}  
\\
\Delta_{n>2}(R) &\sim &
  \begin{cases}
    \frac{1}{R^{n-2}}, &(Rm_\chi \ll 1) \\
\frac{e^{-Rm_\chi}}{R^{n-2}}, &(Rm_\chi \gg 1)
\end{cases}
\end{eqnarray}
We introduce a lepton field $l(x)$ and a Higgs scalar field $h(x)$
localized on our brane. They can interact with the messenger and the
interaction is given by the following Lagrangian \cite{add5}:
\begin{eqnarray}
  M_*^{n-1} {\cal L}^{int} \sim 
  &&\int d^4x'\ \langle \phi \rangle \chi(x',y_*) + \nonumber \\
  &&\int d^4x \ \alpha [l(x)h^*(x)]^2 \chi(x,0),
\label{S}
\end{eqnarray}
where the first part represents the lepton number violation, occurring
on the other brane, and the second part is responsible for the
interaction between SM fields and the messenger on our brane. Let us
assume for simplicity, that the second brane is as far away from ours as
possible, i.e. the distance between the branes is approximately equal to
$R$, the compactification radius. After spontaneous symmetry breaking we
substitute (\ref{chi}) into (\ref{S}), write the Higgs field in terms of
its vev $v$, and identify $l$ with $\nu_L$. We arrive at a mass term of
the Majorana form $ m_{Maj}\ \nu_L^T \ \nu_L $ with the mass given
approximately by \cite{add5}:
\begin{equation}
  m_{Maj} \sim \frac{v^2 \Delta(r)}{M_*^{n-1}}.
\label{mass}
\end{equation}
which, using Eq. (\ref{rf}), may be rewritten as
\cite{mg-extradim1}:
\begin{equation}
  m_{Maj} \sim v^2 R^\frac{n(n-1)}{n+2} M_{Pl}^\frac{2(1-n)}{n+2}
  \Delta_n(R),
\label{mmaj}
\end{equation}
with $ v^2 =(174 \GeV)^2 \sim 10^{22} \eV^2 $ being the Higgs boson vev.

\section{Neutrinoless double beta decay}

Neutrinoless double beta decay ($0\nu 2\beta$) is a process in which a
nucleus undergoes two simultaneous beta decays without emission of
neutrinos
\begin{equation}
  A(Z,N) \to A(Z+2,N-2) + 2e^-.
\end{equation}
It requires neutrino to be a Majorana particle, so that two neutrinos
emitted in beta decays annihilate with each other. It is readily seen
that this process violates the lepton number by two units, thus it is
forbidden in the framework of SM. The $0\nu 2\beta$ decay has been
claimed to be observed \cite{0nu2beta}, but this information has not yet
been confirmed and has met some strong criticism from the community (see
e.g. \cite{0n2b-critic,0n2b-critic2}). In any case, even the limit for
non-observability of this decay sets valuable constraints on the shape
of physics beyond the SM. It is the main tool to verify the nature of
neutrino (Dirac or Majorana particle). It may also help in determining
the absolute value of the neutrino mass.

Ignoring the contributions from right-handed weak currents, the
half-life of $0\nu 2\beta$ can be written in the form \cite{doi}:
\begin{equation}
 T_{1/2} = {\mathcal M}_{spec}^{-1}
  \frac{m_e^2}{\langle m_\nu \rangle^2}.
\label{t12-th}
\end{equation}
One can write the same for any performed or planned $0\nu2\beta$
experiment, replacing the relevant values with the sensitivities of the
experiments
\begin{equation}
  T_{1/2}^{expt.} = {\mathcal M}_{spec}^{-1}
  \frac{m_e^2}{\langle m_\nu \rangle^2_{expt.}},
\label{t12-exp}
\end{equation}
In relations (\ref{t12-th}) and (\ref{t12-exp}) ${\mathcal M}_{spec}$ is
the nuclear matrix element for a specific nucleus which can be
calculated within certain nuclear models, and $m_e$ is the electron
mass. The so-called effective neutrino mass $\langle m_\nu \rangle$ is
defined by the relation
\begin{eqnarray}
\langle m_\nu \rangle &=& \left | \sum_i m_i U_{ei}^2 \right | \nonumber \\
       &\Rightarrow& \left | m_1 \left |U_{e1}\right |^2 \pm m_2 \left
       |U_{e2}\right |^2 \pm m_3 \left | U_{e3} \right |^2 \right |,
\label{mnu}
\end{eqnarray}
where $U$ is the neutrino mixing matrix and $m_i$ are neutrino mass
eigenvalues. The last formulation assumes CP invariance. One sees from
this equation that it is possible to identify $\langle m_\nu \rangle$
with the $ee$ entry of neutrino mass matrix in the flavor basis
\begin{equation}
  \langle m_\nu \rangle = m_{ee},
\label{mnu-mee}
\end{equation}
which is given exactly by the superposition of mass eigenvalues from
Eq. (\ref{mnu}). 

In the next section the link between $T_{1/2}$ and the parameters
describing extra dimensions will be established.

\section{Results}
%
\begin{table}
\caption{\label{tab:0n2b} Sensitivities of some $0\nu2\beta$ experiments
  planned for the future.  The half-life $T_{1/2}^{expt.}$ is expressed in
  years, while effective neutrino mass in eV; $\xi = T_{1/2}^{expt.}\langle
  m_\nu\rangle^2_{expt.}$. Values taken from
  Refs. \cite{vogel,ishihara,gratta,DAMA,EXO}.}
\begin{ruledtabular}
\begin{tabular}{lccc}
            & $T_{1/2}^{expt.}$ & $\langle m_\nu\rangle_{expt.}$ & $\xi$ \\
            \it{expt.} & [y] & [eV] & [y $\eV^2$]  \\
\hline
 DAMA ($^{136}$Xe)   & $1.2 \times 10^{24}$ & 2.3 & $6.3 \times 10^{24}$
\\
 MAJORANA ($^{76}$Ge)& $3 \times 10^{27}$ & 0.044 & $5.8 \times 10^{24}$
\\
 EXO 10t ($^{136}$Xe)& $4 \times 10^{28}$ & 0.012 & $5.7 \times 10^{24}$
\\
 GEM ($^{76}$Ge)     & $7 \times 10^{27}$ & 0.028 & $5.5 \times 10^{24}$
\\
 GENIUS ($^{76}$Ge)  & $1 \times 10^{28}$ & 0.023 & $5.3 \times 10^{24}$
\\
 CANDLES ($^{48}$Ca) & $1 \times 10^{26}$ & 0.2   & $4.0 \times 10^{24}$
\\
 MOON ($^{100}$Mo)   & $1 \times 10^{27}$ & 0.058 & $3.4 \times 10^{24}$
\\
 XMASS ($^{136}$Xe)  & $3 \times 10^{26}$ & 0.10  & $3.0 \times 10^{24}$
\\
 CUORE ($^{130}$Te)  & $2 \times 10^{26}$ & 0.10  & $2.0 \times 10^{24}$
\\
 COBRA ($^{116}$Cd)  & $1 \times 10^{24}$ & 1     & $1.0 \times 10^{24}$
\\
 DCBA ($^{100}$Mo)   & $2 \times 10^{26}$ & 0.07  & $9.8 \times 10^{23}$
\\
 DCBA ($^{82}$Se)    & $3 \times 10^{26}$ & 0.04  & $4.8 \times 10^{23}$
\\
 CAMEO ($^{116}$Cd)  & $1 \times 10^{27}$ & 0.02  & $4.0 \times 10^{23}$
\\
 DCBA ($^{150}$Nd)   & $1 \times 10^{26}$ & 0.02  & $4.0 \times 10^{22}$\\
\end{tabular}
\end{ruledtabular}
\end{table}
%
In the calculations we have neglected the contribution coming from the
third neutrino in the mass basis, setting $|U_{e3}|^2 m_3=0$. This is
justified by the results of CHOOZ experiment \cite{chooz1,chooz2}, which
showed that there is a negligible admixture of $\nu_e$ in the third mass
eigenstate (less than 3\%; most of the analyses place the value of
$|U_{e3}|^2$ between 0 and 0.05). The remaining two mass eigenstates can
be treated as nearly degenerate, with $m_1 \approx m_2$
\cite{pas-weiler}. Within these approximations one may express $m_{ee}$
as a function of $m_1$ and two mixing angles in the following form
\cite{pas-weiler}:
\begin{equation}
  m_{ee}^2 = [1-\sin^2(2\theta_{Solar})\sin^2(\phi_{12}/2)] m_1^2
  \equiv \kappa^{-1} m_1^2,
\label{mee}
\end{equation}
where $\sin^2(2\theta_{Solar}) \approx 0.82$; this value takes into
account the KamLand results \cite{SNO,SNO2,SNO3,KamLand}. By CP symmetry
conservation, the relative phase factor $\phi_{12}$ takes values either
0 or $\pi/2$. For $\phi_{12}=0$, $m_{ee} = m_1$. The more interesting
case involves the mixing angle $\theta_{Solar}$ therefore we chose the
value $\phi_{12}=\pi/2$ having
\begin{equation}
\label{kappa}
  m_{ee}^2 = 1.69 \ m_1^2,
\end{equation}
In fact the exact value of $\kappa$ is irrelevant in our discussion and
the whole coefficient can in principle be set to one. What we are
interested in is the order of magnitude of the half-life of $0\nu2\beta$
decay.

By rewriting $T_{1/2}$ in terms of $T_{1/2}^{expt.}$ we get rid of the
unwanted nuclear matrix elements
\begin{equation}
  T_{1/2} = T_{1/2}^{expt.} 
  \frac{\langle m_\nu \rangle^2_{expt.}}{\langle m_\nu \rangle^2} = 
  \xi \langle m_\nu \rangle^{-2},
\end{equation}
where we have gathered all experimental values in the parameter $\xi =
T_{1/2}^{expt.}\langle m_\nu\rangle^2_{expt.}$.  Eqs. (\ref{mnu-mee})
and (\ref{kappa}) allow to rewrite this expression as
\begin{equation}
  T_{1/2} = \xi m_{ee}^{-2} = \xi \kappa m_1^{-2}.
\end{equation}
Finally, we assume that neutrinos are Majorana particles in order to
discuss the neutrinoless double beta decay. Since we are left with only
one independent mass, we may replace $m_1$ with the expression for
Majorana neutrino mass, Eq.  (\ref{mmaj}), finishing with
\begin{eqnarray}
  T_{1/2} &=& \kappa \ \xi \ v^{-4} \ R^{2n(1-n)/n+2} \nonumber \\
  &\times& M_{Pl}^{4(n-1)/n+2} \ [\Delta_n(R)]^{-2}.
\label{t12}
\end{eqnarray}

One sees that Eq.~(\ref{t12}) consists of two parts. The first one is
connected with $0\nu2\beta$ experiments, for which the relevant values
are presented in Tab.~\ref{tab:0n2b}. We have included tha values as
given by various collaboration assuming, that these represent the best
sensitivity of the projects. Therefore the effective neutrino mass,
which depends on the nuclear matrix elements, may be different for the
same isotope. However the calculations of the matrix elements using
various approaches give on average a~change in $\langle m_\nu \rangle$
200--300\% which is well below the level of accuracy of our discussion.

The second part of Eq.~(\ref{t12}) contains parameters of the extra
dimensions. At this point we need to include constraints coming from
other sources, like supernova and neutron star data \cite{hr1,hr2,hr3,
  hanhart1,hanhart2}, and cosmological models \cite{hall,hannestad}.
Altogether, one of the most complete limits have been derived in Ref.~
\cite{hr3} and read:
\begin{eqnarray}
R < 1.5 \times 10^{-7}\mm  \qquad & \hbox{for}\ n=2, \label{r1} \\
R < 2.6 \times 10^{-9}\mm  \qquad & \hbox{for}\ n=3, \label{r2} \\
R < 3.4 \times 10^{-10}\mm \qquad & \hbox{for}\ n=4. \label{r3}
\end{eqnarray}
Taking into account bounds Eqs.~(\ref{r1})--(\ref{r3}) the formula for
$T_{1/2}$ becomes an inequality. Explicitly Eq. (\ref{t12}) takes the
following forms of \emph{lower bounds} on the half-life of $0\nu2\beta$: \\
Case $n=2$
\begin{eqnarray}
T_{1/2} > 1.11 \times 10^{-15}\ \xi' \ (\log(Rm_\chi))^{-2} &,&
\quad (Rm_\chi \ll 1); \nonumber \\
T_{1/2} > 1.11 \times 10^{-15} \ \xi' \ e^{2Rm_\chi} \ Rm_\chi &,&
\quad (Rm_\chi \gg 1). \nonumber
\end{eqnarray}
Case $n=3$
\begin{eqnarray}
T_{1/2} > 0.039 \ \xi' &,&
\qquad (Rm_\chi \ll 1); \nonumber \\
T_{1/2} > 0.039 \ \xi' \ e^{2Rm_\chi} &,&
\qquad (Rm_\chi \gg 1). \nonumber
\end{eqnarray}
Case $n=4$
\begin{eqnarray}
T_{1/2} > 1.69 \times 10^{11} \ \xi' &,&
\qquad (Rm_\chi \ll 1); \nonumber \\
T_{1/2} > 1.69 \times 10^{11} \ \xi' \ e^{2Rm_\chi} &.&
\qquad (Rm_\chi \gg 1). \nonumber
\end{eqnarray}
We have denoted $\xi'=\xi/(1\eV^2)$. In the above calculations the value
$\kappa=1.69$ has been used, which is consistent with the KamLand
results \cite{KamLand}. The inequalities represent lower bounds on the
half-life. The true value of $T_{1/2}$ may be much bigger than the
bounds itself. We would like to stress, that the following discussion is
valid only under our assumptions, i.e. we live in a brane world and
generate neutrino masses according to the ADD suggestion. If not, our
bounds on $T_{1/2}$ are not valid. Inserting $\xi$ corresponding to
various experiments one obtains specific values of $T_{1/2}$. If
a~positive signal will be recorded by that collaboration, it may be
compared with our values to estimate the values of $n$, $R$ and
$Rm_\chi$.

%
\begin{figure*}
\hbox{
\includegraphics{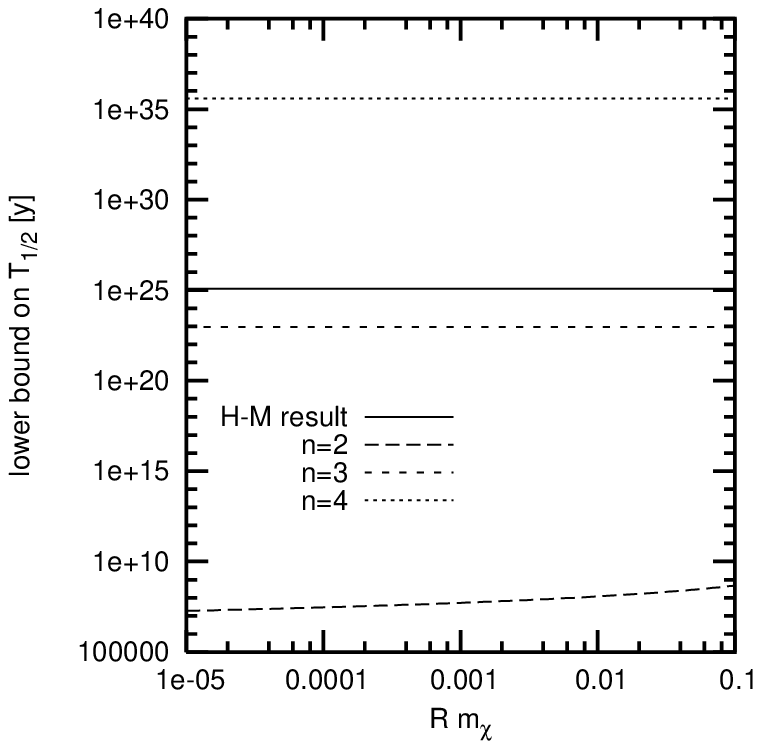}
\includegraphics{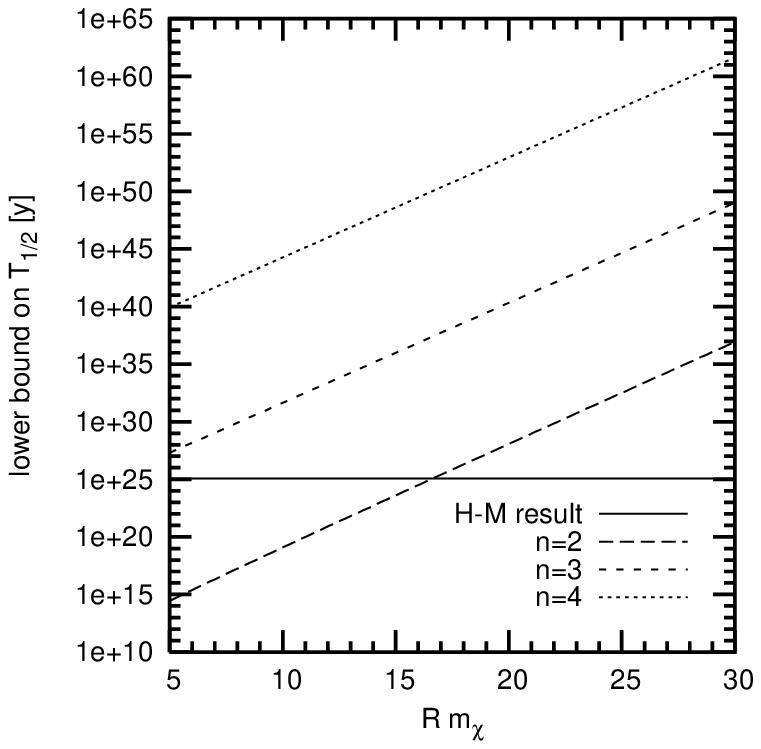}}
\caption{\label{fig:hm} Lower bounds on the half-life of neutrinoless
double beta decay in the case of light (left panel) and heavy (right
panel) messenger particle for the Heidelberg--Moscow experiment.}
\end{figure*}
%
%
\begin{figure*}
\hbox{
\includegraphics{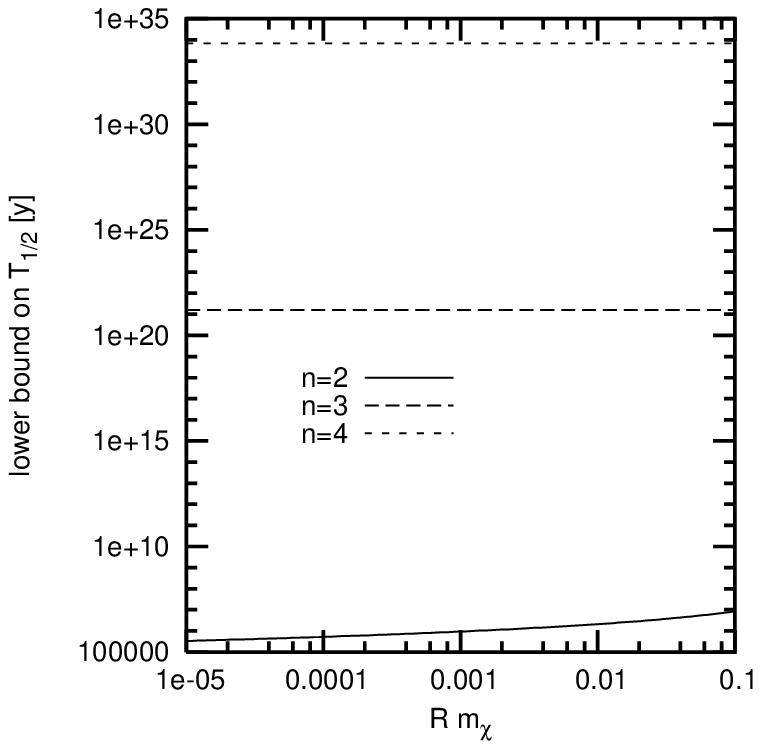}
\includegraphics{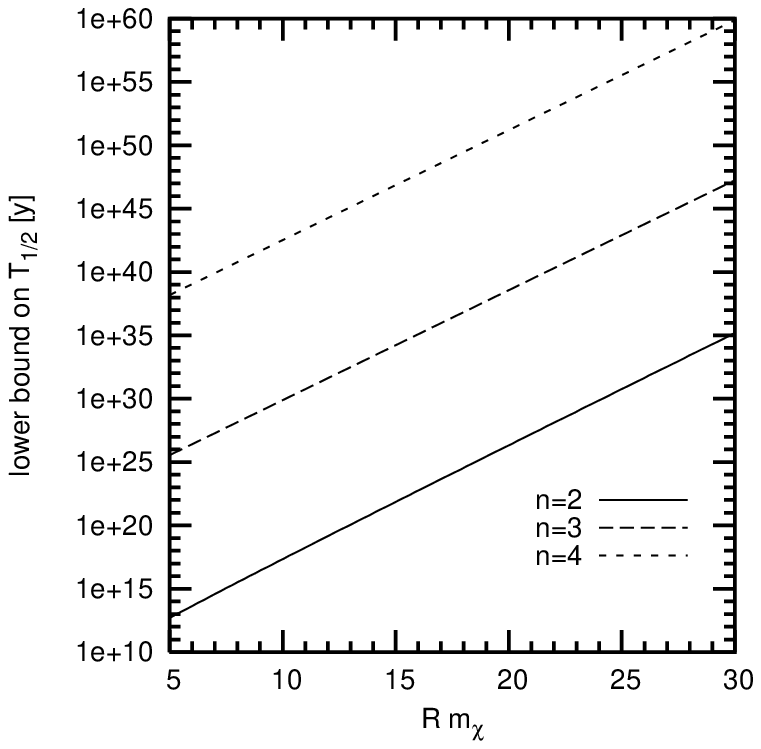}}
\caption{\label{fig:dcba} Lower bounds on the half-life of neutrinoless
double beta decay in the case of light (left panel) and heavy (right
panel) messenger particle for the planned DCBA ($^{150}$Nd) experiment.}
\end{figure*}
%
%
\begin{figure*}
\hbox{
\includegraphics{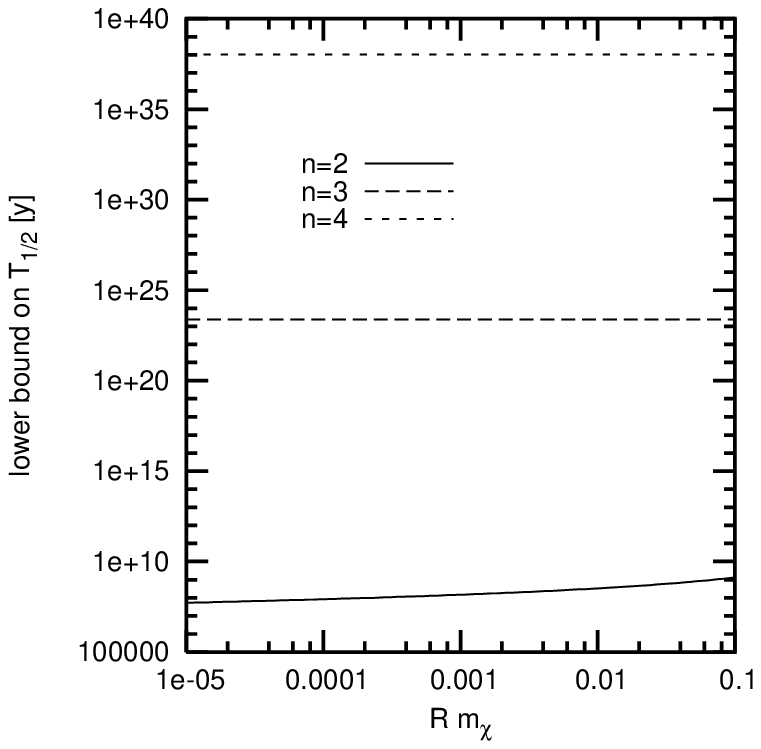}
\includegraphics{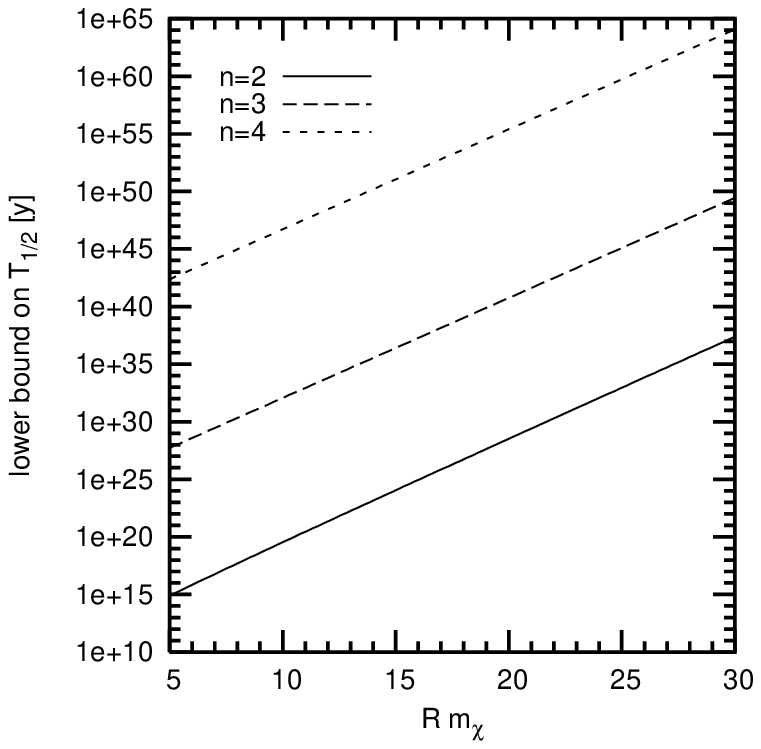}}
\caption{\label{fig:dama} Lower bounds on the half-life of neutrinoless
double beta decay in the case of light (left panel) and heavy (right
panel) messenger particle for the planned DAMA experiment.}
\end{figure*}
%

Let us start with the discussion of the claim of evidence of
$0\nu2\beta$ decay. Klapdor {\it et. al.} in Ref.\ \cite{0nu2beta}
reported a positive signal for the $0\nu2\beta$ transition in $^{76}$Ge
and, after reanalyzing the data, announced the following values at more
than 99.9\% confidence level \cite{klapdor2004}:
\begin{equation}
T_{1/2}^{H-M} = 1.19 \times 10^{25} \y, \qquad
 \langle m_\nu \rangle_{H-M} = 0.44 \eV,
\end{equation}
which corresponds to $\xi = 2.3 \times 10^{24} \y \eV^2$. This claim,
although not confirmed, has not been withdrawn, so it deserves a careful
analysis. The results are presented in Fig.\ \ref{fig:hm}.

One sees that for a light messenger particle, the results do not depend
on $Rm_\chi$, except for $n=2$, where the dependence is very weak. In
fact they are even independent of $R$ and $m_\chi$ separately.
Obtaining constant values for that case (for $n\ge 3$) is a general
feature of the model, as will be seen in further discussion. Assuming
that the H--M values correctly describe $0\nu2\beta$ decay, and taking
into account possible errors, the closest match will be $n=3$. The
$n=2$ case is practically ruled out, since the sensitivities of the
experiments exceeded the obtained value by more than 15 orders of
magnitude, and all of them reported negative results. Another possible
scenario is the $n=4$, but its verification will have to wait for more
powerful experiments than those planned today.

In the case of heavy messenger, the first solution is $n=2$ and
$Rm_\chi\approx 17$. If the H--M results were confirmed, this
possibility is promising. Another choice could be $n=3$ and $Rm_\chi$
being close to 1. This, however, puts a question mark on our
prediction, since we have used the asymptotic al forms of the propagator,
which are not valid for $Rm_\chi\approx 1$. More extra dimensions blow
the half-life to extremely huge values, making the discussion at this
stage meaningless.

\begin{table*}
\caption{\label{tab:n=2} Lower bounds on $T_{1/2}$ in years for two
  additional dimensions ($R<1.5 \times 10^{-7} \mm$).}
\begin{ruledtabular}
\begin{tabular}{lccccccc}
      $Rm_\chi=$ & 0.001 & 0.01 & 0.1 & 10 & 15 & 20 & 25 \\
\hline
 DAMA
& $1.6 \times 10^{8}$ & $3.6 \times 10^{8}$ & $1.4 \times 10^{9}$ & $3.7 \times 10^{19}$ & $1.2 
\times 10^{24}$ & $3.6 \times 10^{28}$ & $1.0 \times 10^{33}$ \\
MAJORANA
& $1.5 \times 10^{8}$ & $3.3 \times 10^{8}$ & $1.3 \times 10^{9}$ & $3.4 \times 10^{19}$ & $1.1 
\times 10^{24}$ & $3.3 \times 10^{28}$ & $9.2 \times 10^{32}$ \\
EXO (10t)
& $1.4 \times 10^{8}$ & $3.2 \times 10^{8}$ & $1.3 \times 10^{9}$ & $3.3 \times 10^{19}$ & $1.1 
\times 10^{24}$ & $3.2 \times 10^{28}$ & $9.0 \times 10^{32}$ \\
GEM
& $1.3 \times 10^{8}$ & $3.1 \times 10^{8}$ & $1.2 \times 10^{9}$ & $3.2 \times 10^{19}$ & $1.0 
\times 10^{24}$ & $3.1 \times 10^{28}$ & $8.6 \times 10^{32}$ \\
GENIUS
& $1.3 \times 10^{8}$ & $3.0 \times 10^{8}$ & $1.2 \times 10^{9}$ & $3.1 \times 10^{19}$ & $1.0 
\times 10^{24}$ & $3.0 \times 10^{28}$ & $8.3 \times 10^{32}$ \\
CANDLES
& $9.7 \times 10^{7}$ & $2.2 \times 10^{8}$ & $8.7 \times 10^{8}$ & $2.3 \times 10^{19}$ & $7.5 
\times 10^{23}$ & $2.2 \times 10^{28}$ & $6.2 \times 10^{32}$ \\
MOON
& $8.2 \times 10^{7}$ & $1.9 \times 10^{8}$ & $7.4 \times 10^{8}$ & $2.0 \times 10^{19}$ & $6.3 
\times 10^{23}$ & $1.9 \times 10^{28}$ & $5.3 \times 10^{32}$ \\
XMASS
& $7.2 \times 10^{7}$ & $1.7 \times 10^{8}$ & $6.5 \times 10^{8}$ & $1.7 \times 10^{19}$ & $5.6 
\times 10^{23}$ & $1.7 \times 10^{28}$ & $4.7 \times 10^{32}$ \\
CUORE
& $4.7 \times 10^{7}$ & $1.1 \times 10^{8}$ & $4.3 \times 10^{8}$ & $1.1 \times 10^{19}$ & $3.7 
\times 10^{23}$ & $1.1 \times 10^{28}$ & $3.1 \times 10^{32}$ \\
COBRA
& $2.4 \times 10^{7}$ & $5.5 \times 10^{7}$ & $2.1 \times 10^{8}$ & $5.7 \times 10^{18}$ & $1.8 
\times 10^{23}$ & $5.5 \times 10^{27}$ & $1.5 \times 10^{32}$ \\
DCBA ($^{100}$Mo)
& $2.3 \times 10^{7}$ & $5.4 \times 10^{7}$ & $2.1 \times 10^{8}$ & $5.6 \times 10^{18}$ & $1.8 
\times 10^{23}$ & $5.4 \times 10^{27}$ & $1.5 \times 10^{32}$ \\
DCBA ($^{82}$Se)
& $1.1 \times 10^{7}$ & $2.6 \times 10^{7}$ & $1.0 \times 10^{8}$ & $2.7 \times 10^{18}$ & $8.8 
\times 10^{22}$ & $2.6 \times 10^{27}$ & $7.4 \times 10^{31}$ \\
CAMEO
& $9.5 \times 10^{6}$ & $2.2 \times 10^{7}$ & $8.5 \times 10^{7}$ & $2.3 \times 10^{18}$ & $7.3 
\times 10^{22}$ & $2.2 \times 10^{27}$ & $6.1 \times 10^{31}$ \\
DCBA ($^{150}$Nd)
& $9.5 \times 10^{5}$ & $2.2 \times 10^{6}$ & $8.5 \times 10^{6}$ & $2.3 \times 10^{17}$ & $7.3 
\times 10^{21}$ & $2.2 \times 10^{26}$ & $6.1 \times 10^{30}$ \\
\end{tabular}
\end{ruledtabular}
\end{table*}
%
\begin{table*}
\caption{\label{tab:n=3} Lower bounds on $T_{1/2}$ in years for three
  additional dimensions ($R<2.6 \times 10^{-9} \mm$). The
  case of light $m_\chi$ is discussed in the text.}
\begin{ruledtabular}
\begin{tabular}{lcccc}
      $Rm_\chi=$ & 3 & 5 & 10 & 15 \\
\hline
DAMA              & $1.1 \times 10^{26}$ & $6.0 \times 10^{27}$ & $1.3 \times 10^{32}$ & $2.9 \times 10^{36}$ \\
MAJORANA          & $1.0 \times 10^{26}$ & $5.5 \times 10^{27}$ & $1.2 \times 10^{32}$ & $2.7 \times 10^{36}$ \\
EXO (10t)         & $1.0 \times 10^{26}$ & $5.4 \times 10^{27}$ & $1.2 \times 10^{32}$ & $2.6 \times 10^{36}$ \\
GEM               & $9.5 \times 10^{25}$ & $5.2 \times 10^{27}$ & $1.1 \times 10^{32}$ & $2.5 \times 10^{36}$ \\
GENIUS            & $9.1 \times 10^{25}$ & $5.0 \times 10^{27}$ & $1.1 \times 10^{32}$ & $2.4 \times 10^{36}$ \\
CANDLES           & $6.8 \times 10^{25}$ & $3.7 \times 10^{27}$ & $8.1 \times 10^{31}$ & $1.8 \times 10^{36}$ \\
MOON              & $5.8 \times 10^{25}$ & $3.2 \times 10^{27}$ & $6.9 \times 10^{31}$ & $1.5 \times 10^{36}$ \\
XMASS             & $5.1 \times 10^{25}$ & $2.8 \times 10^{27}$ & $6.0 \times 10^{31}$ & $1.3 \times 10^{36}$ \\
CUORE             & $3.4 \times 10^{25}$ & $1.8 \times 10^{27}$ & $4.0 \times 10^{31}$ & $9.0 \times 10^{35}$ \\
COBRA             & $1.7 \times 10^{25}$ & $9.2 \times 10^{26}$ & $2.0 \times 10^{31}$ & $4.5 \times 10^{35}$ \\
DCBA ($^{100}$Mo) & $1.6 \times 10^{25}$ & $9.0 \times 10^{26}$ & $2.0 \times 10^{31}$ & $4.4 \times 10^{35}$ \\
DCBA ($^{82}$Se)  & $8.1 \times 10^{24}$ & $4.4 \times 10^{26}$ & $9.6 \times 10^{30}$ & $2.1 \times 10^{35}$ \\
CAMEO             & $6.7 \times 10^{24}$ & $3.6 \times 10^{26}$ & $8.0 \times 10^{30}$ & $1.8 \times 10^{35}$ \\
DCBA ($^{150}$Nd) & $6.7 \times 10^{23}$ & $3.6 \times 10^{25}$ & $8.0 \times 10^{29}$ & $1.8 \times 10^{34}$ \\
\end{tabular}
\end{ruledtabular}
\end{table*}
%

The H--M data's $\xi$ value places itself in the middle of considered
projects.  Let us now say a few words about the currently planned
$0\nu2\beta$ experiments.  The results for the DCBA experiment, which
has the smallest $\xi$ value, and for DAMA experiments with the biggest
$\xi$ are depicted on Figs.\ \ref{fig:dcba} and \ref{fig:dama},
respectively.  For the remaining projects the results are summarized in
Tabs.\ \ref{tab:n=2} and \ref{tab:n=3}. All the values presented in the
tables are bounds from below on the $0\nu2\beta$ half-life. In any case
one should always bear in mind the upper limit on $R$ and from the value
of $Rm_\chi$ deduce the appropriate mass of the messenger particle.

The results for $n=2$ are summarized in Tab.\ \ref{tab:n=2}. The case
of light messenger is excluded by the negative results of $0\nu2\beta$
experiments, which sensitivities have exceeded the given threshold by
more than 15 orders of magnitude. In the case of heavy messenger, the
quantity $Rm_\chi$ turns out to be equal to at least 15. This value
matches well the $Rm_\chi\approx 17$ obtained for H--M experiment.

Case $n=3$ is presented in Tab.\ \ref{tab:n=3} for heavy messenger. One
sees that this case is not limited by the experiments. In fact the
values $Rm_\chi$ around 3 correspond to the sensitivities of currently
performed experiments, like the Germanium neutrinoless double beta decay
in Gran Sasso. We repeat here, that for such small value, the validity
of the results may be questioned. For light $\chi$ particle the
calculations give $T_{1/2} > 3.9 \times 10^{-2}\ \xi'$, so there is no
$m_\chi$ dependence. Since the experimental factor is of the order of
$10^{24}$, we end up with $T_{1/2} > 10^{22-23}\y$.

If the number of extra dimensions is four, a simple calculation gives
$T_{1/2} > 10^{38}\y$ for heavy messenger. This value is beyond the
abilities of any currently planned experiment. It may, however, serve
as clue if all of them report negative results. For a light messenger,
we face a similar situation as previously, i.e. the $m_\chi$ dependence
drops out. For four extra dimensions we finish with $T_{1/2} > 1.69
\times 10^{11} \ \xi'$, therefore obtaining $T_{1/2} > 10^{35}\y$.

The discussion may be extended for more extra dimensions, if one is
interested in the half-life of $0\nu2\beta$ being longer than $10^{38}$
years.

\section{Summary}

Theories which deal with extra dimensions face one basic problem, namely
the obvious difficulty of verification. There are two main types of
experiments performed nowadays. One are the tabletop gravity
experiments, which test the Newton $1/r^2$ law on very small distances.
The best accuracy of present setups is 0.1 mm \cite{long} which is not
sufficient. The other are based on astronomical observations, mainly
concerning supernovas and black holes
\cite{hr1,hr2,hr3,hanhart1,hanhart2,hall,hannestad}. These, however,
are very difficult and highly model-dependent. The work presented first
in Refs.\ \cite{mg-extradim1} and \cite{mg-extradim2}, and continued
here, shows that it is possible to use the results for exotic nuclear
processes also to constrain the theories with large extra dimensions.
Needless to say, we have much better control over these experiments,
than those aforementioned.

To sum up the results of the present paper, one can say that within the
model used, the current $0\nu2\beta$ experiments have reached the
sensitivity to explore the possibility of two, and maybe three
additional spatial dimensions. The future experiments should be able to
rule out the possibility of three extra dimensions, provided they finish
with negative results. The sensitivity needed to reach the $n=4$
threshold should be at least $\sim 10^{35}\y$, which is 7 to 9 orders of
magnitude better than the currently planned experiments can achieve.

\acknowledgments This work has been supported by the Polish State
Committee for Scientific Research under grants no.~2P03B~071~25 and
1P03B~098~27 and the International Graduiertenkolleg GRK683 by the DFG
(Germany). Two of us (MG, WAK) would like to thank Prof. A. Faessler
for his warm hospitality in T\"ubingen during the Summer 2004.

\bibliography{art7}
\end{document}